\RequirePackage{fix-cm}
\documentclass[smallextended]{svjour3}       
\smartqed  
\usepackage{graphicx}
\usepackage{color}
\begin{document}

\title{$^4$He in Nanoporous Media: 4D XY Quantum Criticality at Finite Temperatures
}


\author{Tomoyuki Tani$^1$         \and
        Yusuke Nago$^1$             \and
        Satoshi Murakawa$^2$       \and
        Keiya Shirahama$^1$
}


\institute{
$^1$              Department pf Physics, Keio University, Yokohama 223-8522, Japan \\
$^2$              Cryogenic Research Center, The University of Tokyo, Bunkyo-ku, Tokyo 113-0032, Japan
}


\maketitle

\begin{abstract}
We review our study of critical phenomena in superfluid $^4$He confined in nanoporous glasses. 
$^4$He in nanoporous media is an ideal ground to survey the quantum phase transition of bosons. 
In the present work, critical phenomena were examined using a newly developed hydrodynamic mechanical resonator. 
The critical exponent of superfluid density $\zeta$ was found to be 1.0, in contrast to 0.67 in bulk $^4$He. 
We also demonstrate that the superfluid density is proportional to $|P - P_{\mathrm c}|^{\zeta _p}$ with $\zeta _p = 1$ at any finite temperatures. 
These are the decisive evidences for the 4D XY criticality, which should have been observed only at 0 K, at finite temperatures. 
We propose a mechanism of the quantum criticality at finite temperatures in terms of phase alignment among the nanoscale localized Bose condensates (LBECs) in nanopores. 
The proposed mechanism are discussed in the consideration of the correlation length compared with the quantum effect.
\end{abstract}

\section{Introduction}
\label{sec_introduction}

Quantum phase transitions (QPTs) have been of great interest in physics of strongly correlated systems\cite{SachdevQPT2011,SondhiRMP1997}. 
$^4$He confined in a nanoporous Gelsil glass exhibits a QPT at 0 K\cite{YamamotoPRL2004,YamamotoPRL2008,ShirahamaJLTP2007,ShirahamaLTP2008,ShirahamaJPSJ2008}.
The superfluid transition is strongly suppressed compared to the case of $^4$He in Vycor and in Aerogel, and the superfluid transition temperature $T_{\rm c}$ decreases as the pressure increases eventually reaching 0 K at a critical pressure $P_{\mathrm c} \sim 3.4$ MPa. 
A theoretical analysis has proposed that the superfluid transition at 0 K obeys the 4D XY criticality, which is originated from $d = 3$ spatial dimensions plus $z = 1$ in imaginary time dimension\cite{EggelPRB2011,EggelPhD}. 
Moreover, it has been suggested that, even at finite temperatures, the superfluid transition seems to be affected by the quantum criticality. 
This is apparently in contradiction to general consideration of QPT, in which at finite temperature the system should undergo a thermal phase transition with a classical criticality, the 3D XY in this case. 
In the previous study using torsional oscillator technique\cite{YamamotoPRL2004}, the critical phenomenon in Gelsil has not been unveiled because of the smeared superfluid response in the vicinity of $T_{\rm c}$. 

To survey the origin of this contradiction, we have performed an experiment of hydrodynamic resonance involving the superflow through nanopores of Gelsil using a newly developed mechanical resonator\cite{TaniJPSJ2021,TaniFullpaper}. 
The critical exponent of superfluid density $\rho_{\mathrm s}$, $\zeta$, which is defined as $\rho_{\mathrm s} \propto (1 - T/T_{\mathrm c})^{\zeta}$, was found to be 1.0 in all pressure range realized in the experiment (0.1 - 2.4 MPa).
The exponent $\zeta$ to be 1 shows that the critical phenomenon is governed by the 4D XY universality class. 
This is the decisive evidence for the quantum criticality at finite temperatures. 
In this paper, we discuss the possible origin of the quantum criticality occurring at finite temperatures\cite{TaniFullpaper}.

\section{Setups}\label{sec_setups}

\begin{figure}[t]
\centering
\includegraphics[width=0.9\textwidth]{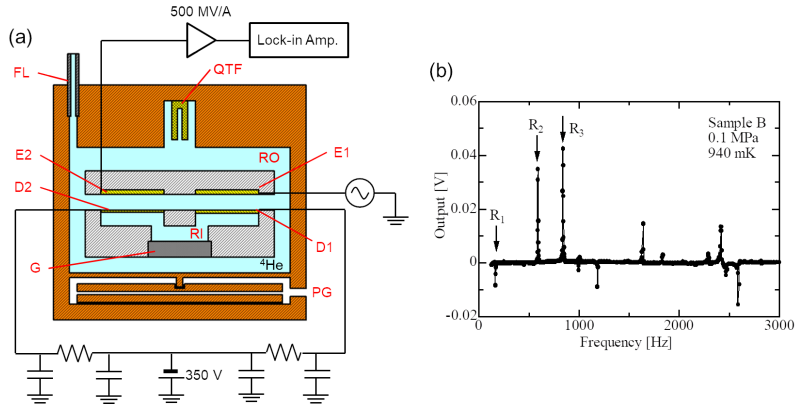}
\caption{(a) A schematic of the setup. Abbreviations are explained in the main text. 
(b) A typical resonance spectrum of the in-phase component of the resonator with sample B. 
}\label{Resonator}
\end{figure}

A schematic illustration of the setup is shown in Fig.~\ref{Resonator}(a). 
Liquid $^4$He is separated into two volumes, inner (RI) and outer reservoir (RO). 
Two Kapton diaphragms act as a part of the wall, and a disk-shaped Gelsil glass (G) hydrodynamically connects RI and RO via nanopores. 
A DC bias voltage (350 V) is applied to two diaphragms. 
The resonance is driven by the AC force exerted on the driver diaphragm (D1), which is induced by an AC voltage (typically 10 V$_{\rm pp}$) applied to the driver electrode (E1). 
The resonant oscillation of D1 propagates to the detector diaphragm (D2) via the motion of bulk liquid $^4$He in the inner reservoir (RI). 
The displacement of D2 is picked up by the detector electrode (E2) as a displacement current from the motion of charged D2, and is measured by a current and a lock-in amplifiers. 
 
The superfluid response of the liquid $^4$He in Gelsil nanopores is involved in the coupled oscillation among two diaphragms and bulk liquid $^4$He in RI. 
A capacitive pressure gauge (PG) and a quartz tuning fork (QTF) are mounted on the copper enclosure to measure pressure and effective viscosity of bulk liquid $^4$He, respectively, simultaneously with the resonance measurement. 

We employed two Gelsil samples from different batches, called sample A and B. 
They have identical pore-size distributions, peaked at about 3 nm\cite{TaniFullpaper}.

\section{Superfluid density}\label{sec_frequency}
\begin{figure}[t]
\centering
\includegraphics[width=1\textwidth]{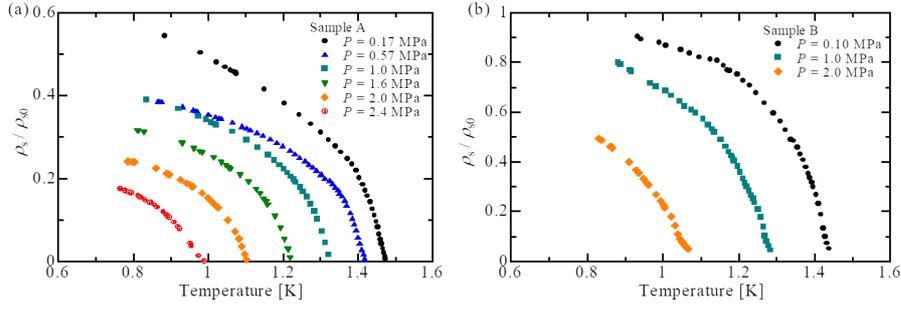}
\caption{Temperature dependencies of the superfluid density $\rho_{\rm s}$ in Gelsil nanopores for (a) sample A and (b) sample B, normalized by $\rho_{\rm s0}$. }\label{SFdensity}
\end{figure}

An example of the frequency spectrum of the resonator is shown in Fig.~\ref{Resonator}(b). 
A number of peaks are observed.  
Only the lowest frequency mode, which is denoted as R$_1$, is attributed purely to the resonant motion of superfluid $^4$He through Gelsil, while all the other modes (R$_2$, R$_3$, $\cdots$) are to coupled oscilllation among two diaphragms and bulk liquid $^4$He involving the superflow through Gelsil, as discussed above. 

A large part of the analysis of the present study is based on the temperature dependencies of the R$_2$ mode.
Although it was difficult to understand completely the mechanism of these modes due to the structural complexity of the resonator, we find that the resonant frequency of the R$_2$ mode, $f_2$, below $T_{\rm c}$ is given by\cite{TaniFullpaper}
\begin{equation}
  f_2 = \frac{1}{2 \pi}\sqrt{\frac{8 \pi (\sigma_1 + \sigma_2)}{\rho V - \alpha \rho_{\rm s}}} , 
\label{Eq_f2}
\end{equation}
where $V$ is the volume of bulk liquid $^4$He, $\rho_{\rm s}$ superfluid density in Gelsil, $\rho$ density of liquid $^4$He, $\sigma_1$ and $\sigma_2$ the tensions of two diaphragms D1 and D2 and $\alpha$ a coefficient determined by hydrodynamic conditions of the resonator. 
The superfluid density $\rho_{\rm s}$ in Gelsil nanopores is calculated from $f_2$. 
The temperature dependencies of $\rho_{\rm s}$ are shown in Fig.~\ref{SFdensity}, in which $\rho_{\rm s}$ is normalized by $\rho_{\rm s0}$, a value obtained by extrapolating the data for the lowest pressure for each sample to zero temperature. 
The details of resonant frequencies and the validations of Eq.~(\ref{Eq_f2}) has been discussed elsewhere\cite{TaniFullpaper}.

\section{Discussions}
\subsection{Critical exponent}\label{sec_ce}

\begin{figure}[t]
\centering
\includegraphics[width=1\textwidth]{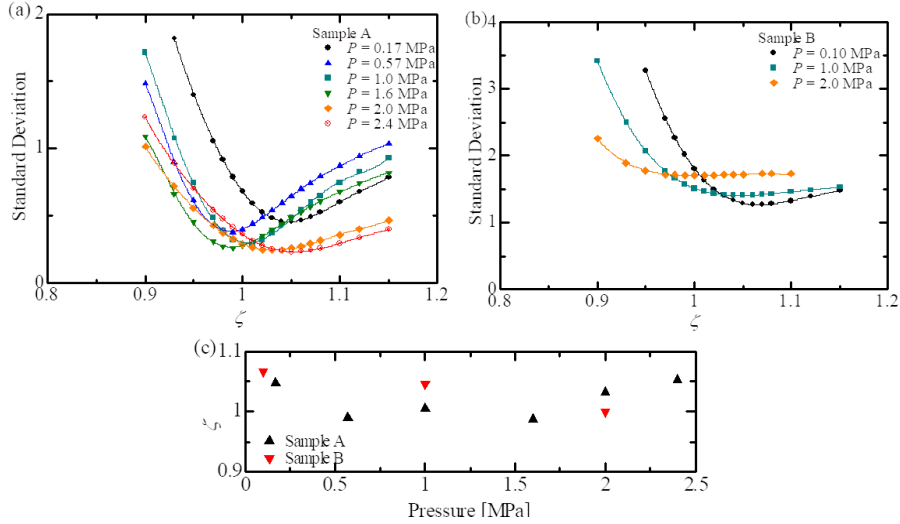}
\caption{The standard deviations of the fittings $\rho_{\rm s} \propto |1 - T/T_{\rm c}|^{\zeta}$ with various $\zeta$ for (a) sample A and (b) B. (c) The critical exponents for each data under various pressures.}\label{StandardDeviation}
\end{figure}

The critical behavior of interest is written by the relation
\begin{equation}
  \rho_{\rm s} \propto |1 - T / T_{\rm c} |^{\zeta}.
\label{Eq_ce}
\end{equation}
Here we precisely determine $\zeta$ by analyzing the standard deviation, the root mean square of residuals, of the fitting by Eq.~(\ref{Eq_ce}) with some different critical exponent $\zeta$, as shown in Figs.~\ref{StandardDeviation}(a) and (b). 
A similar analysis has been adopted in previous study of film $^4$He system on Vycor substrate\cite{BishopPRB1981}.
The best values of $\zeta$ are determined as the value which give the minimum standard deviation. 
The values of $\zeta$ are summarized in Fig.~\ref{StandardDeviation}(c). 
It shows that $\zeta$ is 1 for all the pressures and for two samples. 
Although it is not clear why the standard deviation tends to be asymmetric, 
Figs.~\ref{StandardDeviation}(a) and (b) suggest that $\zeta$ of $^4$He in Gelsil takes obviously a different value from those in other conventional $^4$He systems, especialy $\zeta = 0.67$ in the 3D XY universality class of bulk $^4$He, because the standard deviation sharply grows at lower values of $\zeta$. 
As the superfluid density $\rho_{\mathrm s}$ is related to the superfluid order parameter $\Psi$ by $\rho_{\mathrm s} = \left| \Psi \right|^2$, the critical exponent $\zeta = 1$ corresponds to the critical exponent of the order parameter $\beta$ is 0.5. 
This shows that the $^4$He in Gelsil belongs to the 4D or higher dimensional XY universality class, in which the critical behavior obeys the mean-field theory\cite{NishimoriOxford}. 

\subsection{Critical behavior for pressure and 4D XY universality class}

\begin{figure}
\centering
\includegraphics[width=1\textwidth]{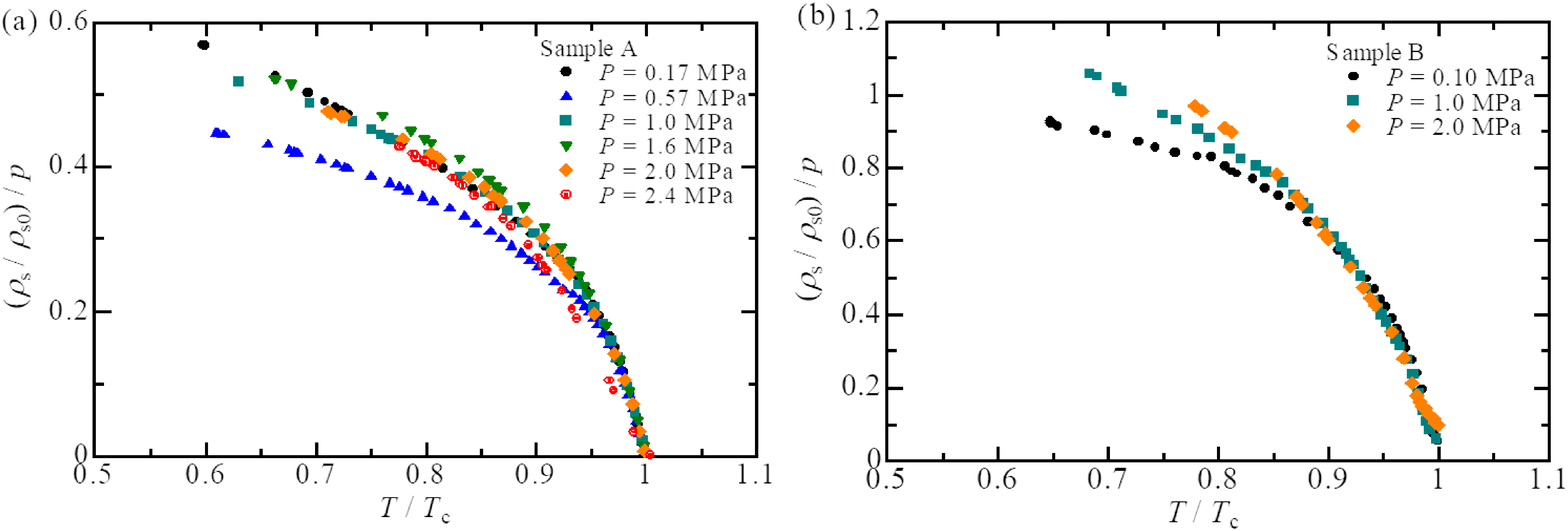}
\caption{$\left(\rho_{\rm s}/\rho_{\mathrm {s0}}\right)/p$ as a function of $T/T_{\rm c}$ for (a) sample A and (b) B.}\label{Scaling}
\end{figure}

Eggel and Oshikawa have suggested that the superfluid $^4$He in Gelsil is governed by 4D XY universality class at least at 0 K in wide pressure range, even sufficiently far away from the quantum critical point\cite{EggelPRB2011,EggelPhD}. 
One of their conclusion is that the superfluid density $\rho_{\rm s}$ is proportional to $| P - P_{\rm c} |^{\zeta_{p}}$, where $\zeta_{p}$ is the critical exponent given by $\zeta_{p} = (d + z - 2)\nu$, which is 1 for 4( $= 3 + 1$)D XY universality class because $\nu = 1/2$ in the mean-field theory. 
This relation should hold at any $T/T_{\rm c}$ of finite temperatures. 
Figure \ref{Scaling} shows that $\rho_{\rm s} / p $ ,where $p = 1 - P/P_{\rm c}$ is reduced pressure, is constant with little deviation at any $T/T_{\rm c}$ for various pressures. 
This demonstrates that $\zeta_{p} = 1$, i.e., again, the system belongs to the 4D XY universality class. 
Therefore, taking into account the discussion in Sec.~\ref{sec_ce}, we conclude that the superfluid $^4$He in Gelsil exhibits 4D XY criticality at finite temperatures.

\subsection{4D quantum criticality at finite temperatures}\label{sec_4DXY}
Eggel and Oshikawa\cite{EggelPRB2011,EggelPhD} explained the 4D XY behavior at 0 K, by applying a disordered Bose-Hubbard model, in which 
nanoscale superfluid droplets (LBEC, see below) are located on regular lattice sites possessing random chemical potential. 
They showed that the system undergoes QPT if the energy increment when one $^4$He atom moves to $i$-th site from its neighbors (a.k.a. the ``charging'' energy in analogue with a Josephson junction array), $V_i$, which indicates the strength of quantum fluctuation, 
is large. 
$V_i$ is given by 
\begin{equation}
  V_i = \left( {\mathcal V}_i \nu ^2 \kappa \right)^{-1},
\end{equation}
where ${\mathcal V}_i$ is volume of $i$-th site, $\nu$ the number density and $\kappa$ the compressibility of $^4$He. 
Using the bulk value for $\kappa$, $V_i$ is estimated to be 0.54 K.
Since $\kappa$ is suppressed by confining into nanopores\cite{GorJCP2015}, $V_i$ can be comparable to $T_{\rm c}$. 
Therefore, the quantum fluctuation can dominate the transition at finite $T$.

\begin{figure}[t]
\centering
\includegraphics[width=0.95\textwidth]{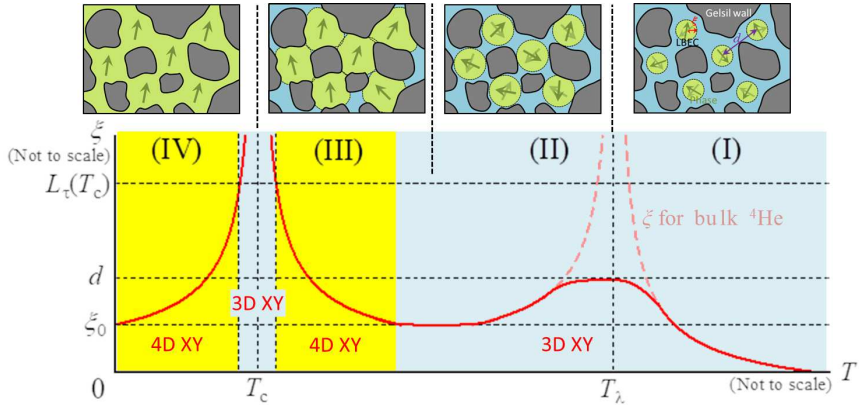}
\caption{Schematical illustrations of growth of LBECs, in which arrows indicate the phase of the LBEC order parameter, and temperature dependence of the correlation length.}\label{LBEC}
\end{figure}

We propose a mechanism explaining the 4D quantum criticality at finite $T$'s. 
Figure~\ref{LBEC}
shows $^4$He in Gelsil at various $T$'s. 
The corresponding change in the correlation length $\xi(T)$ is also schematically shown. 
In the nanopore network of Gelsil, a number of nanoscale superfluid droplets, which we call localized Bose-Einstein condensates (LBECs), grow at and below $T_{\lambda}$ 
(Fig.~\ref{LBEC}
(I) and (II)). 
The existence of LBECs was confirmed by our previous heat capacity study\cite{YamamotoPRL2008}. 
During the formation of  the LBECs, $\xi$ does not diverge but takes a maximum. 
At $T_\lambda > T > T_{\rm c}$, macroscopic superfluidity is \textit{not} realized due to the lack of the phase coherence among LBECs, whose sizes are of the order of the correlation length limited by the size of nanopores. 

As $T$ approaches $T_{\mathrm c}$ from above, the phase coherence grows not only in spatial but in \textit{imaginary time} ($\tau = i\hbar/k_{\mathrm B}T$) dimensions. 
The system eventually undergoes macroscopic superfluid transition at $T_{\mathrm c}$, at which $\xi$ reaches the mean distance between LBECs and the LBECs overlap.
In the imaginary time dimension, the divergence is suppressed at a length scale of the quantum fluctuation\cite{SondhiRMP1997}, which is given by $L_{\tau} = h c / k_{\rm B} T$, where $h$ is Planck's constant, $k_{\rm B}$ Boltzmann constant and $c$ the velocity of a collective excitation, the phonon velocity in this case. 
$L_{\tau}$ is estimated to be 7.5 nm at 1.5 K, which is larger than mean pore size of Gelsil. 
In the proposed mechanism, the quantum fluctuation is still significant even near $T_{\mathrm c}$ of finite temperature, because $\xi$ at the temperature region where the critical behavior is discussed is comparable to $L_{\tau}$, which leads to 4D XY criticality at finite temperature. 
On the other hand, in the very vicinity of $T_{\mathrm c}$, the system is supposed to show a crossover from 4D to 3D. 
In the present study, however, such signs of the crossover were not found at least the present temprature range. 
This requires a further study as an interesting problem.

The proposed mechanism of the superfluid transition of $^4$He confined in Gelsil is also supported by an analysis of the dissipation energy of the resonator below $T_{\mathrm c}$ discussed elsewhere\cite{TaniFullpaper}

\section{Experimental perspectives}
Interesting questions remain to be solved. 
In this work, the critical phenomenon near 0 K is not examined because it locates at higher pressure than the bulk freezing pressure 2.5 MPa. 
Eggel \textit{et al.} proposed that the effect of disorder alter the universality class near $P_{\mathrm c}$\cite{EggelPRB2011}. 
Experiment using a resonator working in bulk solid $^4$He with piezoelectric pressure driver\cite{MukharskyPRB2009} and SQUID-based displacement sensor\cite{MukharskyJLTP2007} may solve this problem. 

A crossover from 4D to 3D XY is expected at temperatures at which $\xi$ $L_{\tau}$.
This possibility was proposed in underdoped cuprates\cite{FranzPRL2006}. 
The 4D-3D crossover will result in the change of $\zeta = 1$ to other value such as 2/3 (in the case of usual 3D XY). 
Further studies in the very vicinity of $T_{\rm c}$ using various Gelsil will reveal the crossover.

\section{Conclusion}
The critical phenomenon of $^4$He in Gelsil was examined using a mechanical resonator. 
At $0.1 < P < 2.4$ MPa, the critical exponent of superfluid density $\zeta$ is found to be 1.0, which is the value of 
the mean-field theory.
The critical behavior of the superfluid density for pressure discussed by Eggel {\it et. al.}\cite{EggelPRB2011,EggelPhD} at zero temperature was found to also hold at any finite temperatures, which means the 4D XY criticality at finite temperatures.
The 4D XY quantum criticality at finite $T$s is explained by a mechanism of the phase coherence among LBECs. 
This work show that $^4$He in nanoporous media is a unique bosonic system exhibiting 4D XY criticality at finite temperatures.

\begin{acknowledgements}
We appreciate fruitful discussions with Kazuyuki Matsumoto and Tomoki Minoguchi. 
\end{acknowledgements}


\begin{thebibliography}{}
\bibitem{SachdevQPT2011}
S. Sachdev, quantum Phase Transitions,Cambridge University Press (2011)
\bibitem{SondhiRMP1997}
S. L. Sondhi, S. M. Girvin, J. P. Carini, and D. Shahar, Rev. Mod. Phys. {\bf 69}, 315 (1997)
\bibitem{YamamotoPRL2004}
K. Yamamoto, H. Nakashima, Y. Shibayama and K. Shirahama, Phys. Rev. Lett. {\bf 93}, 075302 (2004)
\bibitem{YamamotoPRL2008}
K. Yamamoto, Y. Shibayama and K. Shirahama, Phys. Rev. Lett. {\bf 100}, 195301 (2008)
\bibitem{ShirahamaJLTP2007}
K. Shirahama, J. Low Temp. Phys. {\bf 146}, 485-497 (2007)
\bibitem{ShirahamaLTP2008}
K. Shirahama, K. Yamamoto and Y. Shibayama, Low Temp. Phys. {\bf 34}, 273 (2008)
\bibitem{ShirahamaJPSJ2008}
K. Shirahama, K. Yamamoto and Y. Shibayama, J. Phys. Soc. Jpn. {\bf 77}, 111011 (2008)
\bibitem{EggelPRB2011}
Th. Eggel, M. Oshikawa and K. Shirahama, Phys. Rev. B {\bf 84}, 020515(R) (2011)
\bibitem{EggelPhD}
Th. Eggel, Ph.D. Thesis, University of Tokyo (2011)
\bibitem{TaniJPSJ2021}
T. Tani, Y. Nago, S. Murakawa, and K. Shirahama, J. Phys. Soc. Jpn. {\bf 90}, 033601 (2021)
\bibitem{TaniFullpaper}
T. Tani, Y. Nago, S. Murakawa, and K. Shirahama, J. Phys. Soc. Jpn. {\bf 91}, 014603 (2022)
\bibitem{BishopPRB1981}
D. J. Bishop, J. E. Berthold, J. M. Parpia, and J. D. Reppy, Phys. Rev. B {\bf 24}, 5047 (1981)
\bibitem{NishimoriOxford}
H. Nishimori and G. Ortiz, {\it Elements of Phse Transitions and Critical Phenomena} (Oxford University Press, Oxford, U. K. (2021))
\bibitem{GorJCP2015}
G. Y. Gor, D. W. Siderius, C. J. Raumussen, W. P. Krekelberg, V. K. Shen and N. Bernstein, J. Chem. Phys. {\bf 143}, 194506 (2015)
\bibitem{MukharskyPRB2009}
Y. Mukharsky, A. Penzev, and E. Varoquaux, Phys. Rev. B {\bf 80}, 140504(R) (2009)
\bibitem{MukharskyJLTP2007}
Y. Mukharsky, O. Avenel, and E. Varoquaux, J. Low Temp. Phys. {\bf 148}, 689 (2007)
\bibitem{FranzPRL2006}
M. Franz and A. P. Iyengar, Phys. Rev. Lett. {\bf 96}, 047007 (2006)
\end{thebibliography}
\end{document}